\documentstyle[12pt,epsf]{article}

\newcommand{\bmat}{\left(\begin{array}}
\newcommand{\emat}{\end{array}\right)}

\def\yzero{\smash{\hbox{$y\kern-4pt\raise1pt\hbox{${}^\circ$}$}}}

\def\-{\hphantom{-}}

\def\s2{\frac{1}{\sqrt2}}

\def\beq{\begin{equation}}
\def\eeq{\end{equation}}
\def\beqa{\begin{eqnarray}}
\def\eeqa{\end{eqnarray}}

\def\IF{\relax{\rm I\kern-.18em F}}
\def\II{\relax{\rm I\kern-.18em I}}
\def\IP{\relax{\rm I\kern-.18em P}}
\def\inbar{\vrule height1.5ex width.4pt depth0pt}
\def\IC{\relax\hbox{\kern.25em$\inbar\kern-.3em{\rm C}$}}
\def\IR{\relax{\rm I\kern-.18em R}}

\def\Dsl{\,\raise.15ex\hbox{/}\mkern-13.5mu D} 
\def\IZ{Z}
\def\id{{\rm I}}

\topmargin
-1.5cm
\textwidth
15.5cm
\textheight
24cm
\oddsidemargin
0.7cm
\evensidemargin
1.2cm

\begin{document}

\makeatletter
\@addtoreset{equation}{section}
\makeatother
\renewcommand{\theequation}{\thesection.\arabic{equation}}
\pagestyle{empty}
\rightline{FTUAM-01/14, IFT-UAM/CSIC-01-22}
\rightline{\tt hep-th/0107036}
\vspace{0.5cm}
\begin{center}
\LARGE{Branes at angles, torons,\\
stability and supersymmetry\\[10mm]}

\large{
R. Rabad\'an
\\[2mm]
}

\small{
Departamento de F\'{\i}sica Te\'orica C-XI
and Instituto de F\'{\i}sica Te\'orica  C-XVI,\\[-0.3em]
Universidad Aut\'onoma de Madrid,
Cantoblanco, 28049 Madrid, Spain.\\[4mm]
 }
\small{\bf Abstract} \\[7mm]
\end{center}

\begin{center}
\begin{minipage}[h]{14.0cm}

{\small}


We elucidate some properties of the relation between two T-dual systems in tori, branes at angles and branes wrapping the whole torus carrying fluxes. We analyze different features of these systems: charges, low energy spectrum, tadpole cancellation, symmetry groups, ...  and the correspondence between the two viewpoints. Particular attention is paid to supersymmetry and stability conditions.
While on the branes at angles side stability and supersymmetry can be expressed as conditions on the angles between the two branes at the intersection, on the dual side supersymmetry has to do with a correction to Hermite Yang-Mills and a modified notion of stability should be considered.

\end{minipage}
\end{center}
\newpage
\setcounter{page}{1}
\pagestyle{plain}
\renewcommand{\thefootnote}{\arabic{footnote}}
\setcounter{footnote}{0}

\tableofcontents

\newpage

\section{Introduction}

The aim of this paper is to analyze the correspondence between two T-dual systems on tori. On the one hand there are D-branes wrapping submanifolds of half the torus dimension, i.e. middle cycle. These branes intersect generically at a finite number of points. The intersections are specified by some angles. This is what we call the branes at angles viewpoint ( A side of the map). On the other hand, there are  D-branes wrapping the whole torus but carrying non-trivial gauge bundles. These bundles, we will call them torons, are specified by some non-vanishing fluxes ( B side of the map).  Different features obtained from one system can be compared to the other construction. We elaborate the map between these constructions. These toroidal constructions constitute a very simple example to check how mirror symmetry extends to open string systems.

D-branes at angles were first analyzed on flat space  \cite{bdl96,asj96}. The string spectrum, the cylinder partition function, the Ramond-Ramond couplings can be easily obtained. These brane systems can be generalized to toroidal compactifications, orbifolds, orientifolds, ... \cite{oriangles,hepth}. Interestingly enough, these constructions are of a huge phenomenological potential and provide models with a very close structure to the one of the standard model \cite{bgkl00,bkl00,hepph,sm}. 

T-duality in some directions maps generically these branes intersecting at angles to branes wrapping the whole torus and carrying some gauge flux.  There are constructions which are analyzed directly from this dual point of view \cite{b95,aads00}. Some aspects of the  relation between the dual systems, brane at angles and fluxes, in the toroidal case have also been studied in \cite{bgkl00,bkl00}\footnote{See also \cite{cp97}, where this system was analyzed in relation to black holes.}.

This map has been analyzed more deeply in \cite{gr971,gr972,wt,ohta}, in particular, the relation between the D-brane charges. On the side of D-branes wrapping middle cycles, D-branes charges are related to the homology of the cycle. On the other side, D-branes are classified by the K-theory charges, in this case the rank  and the Chern numbers. 


Particular attention is paid in this paper to analyze the correspondence between the stability conditions in the two viewpoints. Let us consider a system of two branes wrapping some submanifold of half of the dimension of the torus ( A side). In this case stability depends on the angles between the two branes at the intersection points. On a two dimensional torus non parallel branes are unstable. This can be seen from a geometrical point of view: there is a submanifold in the same homology class that lowers the volume. From the stringy point of view the instability can be related to the presence of a tachyon at each intersection between the branes. On a four dimensional torus there is a line of stable configurations in the angle space. Angles on  this line indicate that the system preserves some supersymmetry. And in a six dimensional case there is a whole region were the branes do not have tachyons  \cite{km99}. On the side with fluxes, stability conditions can be seen as the stability condition of some bundles. The stability conditions on the branes at angles side can be identified with the $\Pi$-stability of \cite{dfr00,d00}. On the two dimensional torus the stability condition is equivalent to the classical $\mu$-stability. That is not the case for higher tori, where additional terms with higher Chern classes are needed.

A very similar issue happens to supersymmetry conditions. In the two and four dimensional tori, supersymmetry conditions are equivalent to stability conditions, i.e. two branes will be stable (tachyon free from the stringy point of view) if they form a supersymmetric system. This is not the case for the six dimensional torus, where there is a region where branes can be stable without supersymmetry. Supersymmetric conditions on the flux side can be expressed as solutions of an equation that coincides with the one of \cite{mmms99}.

In this paper we will try to clarify these relations. We study the T-dual systems and the correspondence between them: charges (wrapping numbers, K-theoretical charges), spectrum (chiral fermions, tachyonic modes), stability conditions, supersymmetry conditions, ... The following table summarizes some of the relations between the two systems:

\vspace{2mm}

\begin{center}
\begin{tabular}{|c|c|}
\hline
Brane at angles & Torons \\
\hline
\hline
Wrapping numbers & K-theory charges \\
factorizable cycle & $U(1)$ bundle \\
non-factorizable cycle & $SU(N)$ flux \\
supersymmetric cycles & solutions to the MMMS equation \\
stable intersection & stable bundle \\
\hline
\end{tabular}
\end{center}

\vspace{8mm}

We will study two, four and six dimensional tori, although the discussion can be straightforwardly generalized to the eight dimensional case. We consider the transverse space to the torus to be non-compact flat space. We will not be very specific about the Dp-brane we are using in each case, if it is expanding all non-compact dimensions, but only how it wraps the torus. Only for the discussion of the tadpole conditions and chiral fermions this specification will be relevant. In these cases we will consider that the branes expand all non-compact directions.

In section 2, we revisit the construction of branes wrapping middle cycles. In section 3, we analyze the T-dual system with branes carrying fluxes and the relation to the previous construction. In section 4 we study the condition to preserve supersymmetry in both systems and in the last section we analyze stability conditions.

\section{Branes at angles}

\subsection{Generalities}

In this section, we consider branes wrapping middle cycles of different dimension tori ( $A$ branes). These tori will be split into two dimensional tori. The D-branes will wrap straight 1-cycles on each $T^2$. Of course, that is not the most general construction but it allows us to study some of the characteristic features of higher dimensional tori from the knowledge we have of the two dimensional one.

The metric on the two dimensional tori is specified by two complex fields:

\begin{itemize}
\item the complex structure: $U$,
\item the complexified Kahler form: the usual Kahler form is proportional to the area of the two dimensional torus and it is complexified by adding a $B$-field ($B + i A$).
\end{itemize}  

As we will see, some properties of the system depend on the homology of the cycles where the D-branes live. Other properties, as the mass of the fields on the intersection and stability, depend only on the complex structure moduli, and other, as the gauge coupling constants in the effective lower dimensional theory, on the whole moduli. 

The stability of the system is related to the presence of tachyons in the spectrum, i.e. to the complex structure moduli. This property has a geometrical meaning. The geometrical related problem is to find some manifold in the same homology class as the sum of the other two and with a lower volume. Stability, presence of tachyons in our case, can be analyzed from the angles of the two planes at the intersection. Stability is then a local property \cite{d,l95}. If $\Sigma_a$ is the first plane and $\bar{\Sigma}_b$ is the orientation reversal of the second one, the angles between them of the group $SO(2n)$ take eigenvalues $e^{i\theta_i}$. The theorem says that if the following inequalities are satisfied:
\beq
\theta_i \leq \sum_{j \neq i} \theta_j
\eeq
the system of the two planes is area minimizing. Notice that this is a local geometric condition related to the presence of a tachyon in the intersection and as we will see represents the positivity of the mass of the scalar that could become a tachyon.

Starting from the two dimensional torus we will study higher dimensional tori. In the following chapter we will compare these systems to brane with fluxes, that we will denote as torons.

\subsection{Two dimensional torus}

Let us first consider the simplest case: a two dimensional torus. There are two independent 1-cycles, $[a]$ and $[b]$, which constitutes a basis for $H_1(T^2,Z)$. We will consider D-branes wrapping a straight lines on 1-cycle of the form:
\beq
[\Pi] = n [a] + m [b] 
\eeq

Notice that for a fixed homology class there is just one straight line up to translations. Sometimes we will use the word cycle for both the cycle and the minimizing volume submanifold in that cycle class. 

Two D-branes wrapping cycles $[\Pi]_a$ and $[\Pi]_b$ intersect generically at a finite number of points. At each intersection there is a massless chiral fermion. The presence of this chiral fermion does not depend on the metric on the torus. Changing the metric on the torus, in particular its complex structure, will change the mass of all the tower of fields living at the intersection but the net number of chiral fermions will be conserved.

At each intersection there is also a scalar field with a mass:
\beq
m^2 \alpha' = - \frac{\theta}{2 \pi}
\eeq
where $\theta$ is the absolute value of the angle between the two branes at the intersection, defined in the range $0 \leq \theta \leq \pi$. Notice that for every angle different from zero the scalar field is tachyonic, leading to a decay to a lower volume brane that preserves the RR charges, i.e., it is in the same homology class as the sum of the homology classes of the other two. Notice also that the specific value of the angle depends on the complex structure of the two torus, but not on the Kahler one.

\subsection{Four dimensional torus}

In this case we will consider D-branes wrapping 2-cycles, in particular, a very special type of 2-cycles: those that are the product of two 1-cycles in each two dimensional tori. We call these type of cycles factorizable cycles. This is not the most general case as we can consider branes wrapping one of the two tori \footnote{These two 2-cycles complete a basis for $H_2(T^4,Z)$, as $b_2 = 6$. The remaining two cycles are just the first two dimensional torus ($T^2 \times {p_2}$) and the second one (${p_1} \times  T^2$).}.
A basis for the factorizable cycles subspace is: $[a] \times [a]$, $[a] \times [b]$,  $[b] \times [a]$,  $[b] \times [b]$. If the homology of each 1-cycle is specified by two integers $n_i [a]_i + m_i [b]_i$, the coordinates of the cycle on the above basis are:
\beq
\vec{q} = (n_1 n_2, n_1 m_2, m_1 n_2, m_1 m_2)
\eeq  

We call the cycles of the subspace specified by the $q$ elements,  the $q$-basis. Notice that coordinates specified by 1-cycle numbers satisfy the relation:
\beq
q_1 q_4 = q_2 q_3
\eeq

So, not every cycle in the $q$-subspace can be expressed as the product of two 1-cycles. This condition tells us that in order to get a factorizable 2-cycle one should impose the above condition \footnote{This is analogous to vector spaces of the condition for a vector ($v \in C^4$ for example) to be the tensor product of other two ($v = v_1 \otimes v_2$, where $v_i \in C^2$).}.  In the dual toron picture this condition is mapped to the form of $c_2$ in a $U(1)$ bundle. We will later explain this relation in detail.

As noticed in \cite{gr971} this condition indicates that the cycle has no self-intersection. That can be checked by taking  the intersection matrix in this $q$-basis:
\beq
I = 
\left(
\begin{array}{cccc}
0 & 0 & 0 & 1 \\
0 & 0 & -1 & 0 \\
0 & -1 & 0 & 0 \\
1 & 0 & 0 & 0 
\end{array}
\right)
\eeq  

That is, a 2-cycle in a four dimensional torus can be expressed as the product of two 1-cycles (a factorizable 2-cycle) iff its self intersection number vanishes.

Given a general 2-cycle in the $q$-basis, is it always possible to express it as a sum of two non self-intersecting cycles? The answer is positive. Consider the projection matrices:
 \beq
P_a = 
\left(
\begin{array}{cccc}
1 & 0 & 0 & 0 \\
0 & 0 & 0 & 0 \\
0 & 0 & 1 & 0 \\
0 & 0 & 0 & 0 
\end{array}
\right)
\eeq  
and 
\beq
P_b = 
\left(
\begin{array}{cccc}
0 & 0 & 0 & 0 \\
0 & 1 & 0 & 0 \\
0 & 0 & 0 & 0 \\
0 & 0 & 0 & 1 
\end{array}
\right).
\eeq  

We can decompose the cycle $q$ into $q = q_a + q_b$, where $q_a = P_a q$ and $q_b = P_b q$. Notice that due to the properties $P_a^T I P_a = 0$ and $P_b^T I P_b = 0$, the $q_a$ and $q_b$ cycles do not have self-intersection. All the intersections are obtained from  $P_a^T I P_b + P_b^T I P_a$. In the T-dual picture this decomposition means that we will need a $SU(N)$ flux in addition to the $U(1)$ bundle (see \cite{gr971}). This decomposition can be extended to higher tori with exactly the same meaning in the T-dual picture.

The split of the four dimensional torus into two dimensional ones allows us to extract a pair of angles, one for each two dimensional torus: $\theta_1$ and $\theta_2$. In this case there are two light scalars that can be tachyonic. The masses of these scalars are:
\beqa
m_1^2 \alpha' = - \frac{\theta_1 - \theta_2}{2 \pi} \nonumber \\
m_2^2 \alpha' = - \frac{\theta_2 - \theta_1}{2 \pi} 
\eeqa

Notice that the sum of the square of the masses of these scalars vanishes, i.e. if one of the scalars is massive then the other is tachyonic. Only when the two angles are equal the system is tachyon free. In this case the system is also supersymmetric (we will discuss later the supersymmetric conditions).


Now we will make a proposal to extend the notion of angles to arbitrary cycles, factorizable and non-factorizable cycles. To illustrate the idea let us start with a two dimensional torus and the 1-form $\Omega = dx + \tau dy$. Let us put a brane on the cycle $(n,m)$. Integrating the form along the cycle we get a complex number $\omega_q = R(n+m \tau)$ whose modulus is the length and its argument is the angle relative to the $x_1$-axis $\phi_q$. The relation between the angles between two branes $\theta_{ab}$ and the arguments obtained with the above procedure, $\phi_a$ and $\phi_b$, is the following:
\beq
\theta_{ab} = 
\Bigg\{ \begin{array}{c}
|\phi_a -\phi_b| \ if \ 0 \leq |\phi_a -\phi_b| \leq \pi \\
2 \pi - |\phi_a -\phi_b| \ if  \ \pi \leq |\phi_a -\phi_b| \leq 2 \pi
\end{array}
\label{3}
\eeq

Notice that the definition of the angles $\phi$ and $\theta$ depends only on the complex structure moduli of the two dimensional torus and not on the area.

Now we can extend the notion of angles to the $q$ basis in the four dimensional case. First notice that the holomorphic two form $\Omega$ is:
\beq
\Omega = (dx_1 + \tau_1 dy_1)\wedge(dx_2 + \tau_2 dy_2)
\eeq
where $\tau_i$ is the complex structure of the corresponding $T^2$. 

We can integrate the two form along a 2-cycle $q$:
\beq
\omega_q = q_1 + \tau_2 q_2 + \tau_1 q_1 + \tau_1 \tau_2 q_4 
\eeq

Notice that the phase of the complex number $\omega_q$ is the sum of the angles between the 2-cycle and the $(1,0,0,0)$ and that $|\omega_q|$ is the volume of the minimal submanifold in the q-cycle. Expressed in terms of the angles  $\phi_1$ and $\phi_2$, the  argument of the period of the holomorphic two form is:
\beq
Arg (\omega_q) = \phi_1 + \phi_2
\label{1}
\eeq

We will choose $0 \leq \phi_1 + \phi_2 \leq 2 \pi$. Changing one of the angles from $0$ to $\pi$ we are changing branes into antibranes. To fix the angles, we will choose the angles in such a way that the above inequalities are satisfied. Taking a different value of the argument we are reshuffling fields but keeping the whole structure. The angle in another point of the complex structure moduli is defined by analytical continuation \cite{dfr00,d00}.

Similarly, one can define the holomorphic forms:
\beqa
\Omega' = (dx_1 + \bar\tau_1 dy_1)\wedge(dx_2 + \tau_2 dy_2) \nonumber \\
\Omega'' = (dx_1 + \tau_1 dy_1)\wedge(dx_2 + \bar\tau_2 dy_2)
\eeqa

These forms are related to the angles in the following way:
\beqa
Arg (\omega_q') = - \phi_1 + \phi_2 \nonumber \\
Arg (\omega_q'') =  \phi_1 - \phi_2
\label{2}
\eeqa

Now we have to relate this angles to the $\theta$ angles between branes. To do so, we can proceed as in the two dimensional case explained above: from equations (\ref{1}) and (\ref{2}) we can obtain the $\phi$ angles \footnote{There is an ambiguity due to the simultaneous change of orientation in the two planes that does not affect the physical quantities.}, and using the relation (\ref{3}) we can find the $\theta$ angles. Then we can just plug them into the formulae for the masses of the tachyons. The relation between the periods and the tachyon masses is straightforward for factorizable cycles. For the other cycles it remains as a proposal. This arguments can be generalized to higher dimensional tori. Notice also the parallelism between the formula for the angles here conjectured and the notion of $\Pi$-stability in \cite{dfr00}.

\subsection{Six dimensional torus}

Just as in the previous section, we can define a basis for the subspace of the 3-cycles that can be obtained as a product of three 1-cycles, i.e. factorizable cycles. The basis is of the form:

\begin{center}

\begin{tabular}{ccc}
3-cycle & 1-cycles basis & coordinates \\ \hline
$q_1$ & $[a] \times [a] \times [a]$ & $n_1 n_2 n_3$ \\
$q_2$ & $[a] \times [a] \times [b]$ & $n_1 n_2 m_3$ \\
$q_3$ & $[a] \times [b] \times [a]$ & $n_1 m_2 n_3$ \\
$q_4$ & $[a] \times [b] \times [b]$ & $n_1 m_2 m_3$ \\
$q_5$ & $[b] \times [a] \times [a]$ & $m_1 n_2 n_3$ \\
$q_6$ & $[b] \times [a] \times [b]$ & $m_1 n_2 m_3$ \\
$q_7$ & $[b] \times [b] \times [a]$ & $m_1 m_2 n_3$ \\
$q_8$ & $[b] \times [b] \times [b]$ & $m_1 m_2 m_3$ \\
\end{tabular}

\end{center}

Notice that as in the four dimensional torus there are some non-linear relations between the coordinates:
\beqa
q_1 q_8 = q_2 q_7 \nonumber \\
q_1 q_8 = q_3 q_6 \nonumber \\
q_1 q_8 = q_4 q_5 \nonumber \\
q_2 q_3 q_5 = (q_1)^2 q_8
\label{q}
\eeqa

Thus, not every cycle in this q-basis subspace can be expressed as the product of three 1-cycles, i.e. not every cycle in the q-basis will be factorizable. We will see in the next section that the meaning of these relations in the T-dual picture is that we are dealing with a $U(1)$ bundle. Notice that in this case the self-intersection is always zero  \footnote{The intersection matrix is antisymmetric.}, so the above conditions are not related to the self intersection number as in the four dimensional torus.

If the conditions \ref{q} are not satisfied it implies, in the T-dual picture, a $SU(N)$ flux. Equivalently, not all the Chern numbers can be obtained from the first Chern numbers. That is only possible if the above conditions are satisfied. We will see this point in detail in the next section. 

There are four scalar fields living at each intersection that could become tachyonic, with a mass depending on the angles:
\beqa
m^2_1 \alpha' = \frac{1}{2 \pi}(- \theta_1 +\theta_2 +\theta_3) \nonumber \\
m^2_2 \alpha' = \frac{1}{2 \pi}( \theta_1 -\theta_2 +\theta_3) \nonumber \\
m^2_3 \alpha' = \frac{1}{2 \pi}( \theta_1 +\theta_2 -\theta_3) \nonumber \\
m^2_4 \alpha' = 1 - \frac{1}{2 \pi}( \theta_1 +\theta_2 +\theta_3)
\label{tachyons}
\eeqa

For a homology class the submanifolds that minimize the volume are special Lagrangian \cite{bbh00}. Generically, two submanifolds do not minimize the volume so the branes decay to another brane with minimal area in the same homology class as the sum of the classes of the other two branes. In this case, there is a region where there is not a single Lagrangian submanifold in this class that lowers the volume. In this case the system is stable. From a stringy point of view the absence of lowering volume manifolds can be seen as the absence of tachyons in our system.

A general cycle in the q-basis does not need to satisfy the relations (\ref{q}), but as in the four dimensional case, we can split it into a pair of cycles that satisfy it:
\beq
q = q_a + q_b
\eeq
where $q_a = P_a q$ and $q_b = P_b q$, and 
\beq
P_a = 
\left(
\begin{array}{cc}
\id_4 & 0  \\
0 & 0_4 
\end{array}
\right)
\eeq  
\beq
P_b = 
\left(
\begin{array}{cc}
0_4 & 0  \\
0 & \id_4 
\end{array}
\right).
\eeq  

As in the four dimensional torus, the mass of the scalars can be directly obtained from the periods of the following forms:
\beqa
\Omega_1 = (dx_1 + \bar\tau_1 dy_1)\wedge(dx_2 + \tau_2 dy_2)\wedge (dx_3 + \tau_3 dy_3)\nonumber \\
\Omega_2 = (dx_1 + \tau_1 dy_1)\wedge(dx_2 + \bar\tau_2 dy_2)\wedge (dx_3 + \tau_3 dy_3)\nonumber \\
\Omega_3 = (dx_1 + \tau_1 dy_1)\wedge(dx_2 + \tau_2 dy_2) \wedge(dx_3 + \bar\tau_3 dy_3)\nonumber \\
\Omega_4 = (dx_1 + \tau_1 dy_1)\wedge(dx_2 + \tau_2 dy_2)\wedge (dx_3 + \tau_3 dy_3)
\eeqa

By integrating them along the cycles where the branes are, we can obtain the volumes and the $\phi$ angles. Following the procedure explained in the previous section we can get the $\theta$ angles. As in the four dimensional torus the relation between the argument of the periods and the masses of the tachyon are straightforward for the factorizable cycles case. For the other cases, it remains a proposal.

\begin{figure}
\centering
\epsfxsize=3in
\hspace*{0in}\vspace*{.2in}
\epsffile{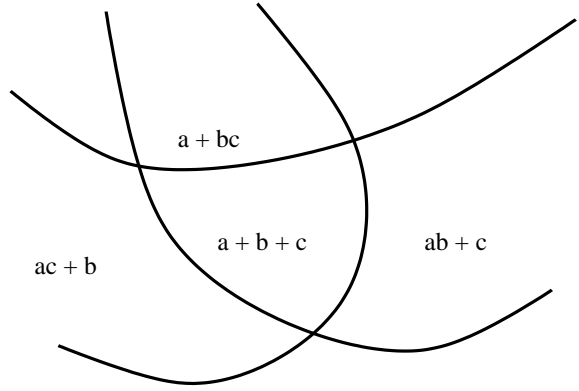}
\caption{\small Schematic draw that represents the moduli space of complex structures. In each of these regions a different number of branes is stable. Lines represent transitions where pairs of branes decay to a more stable brane in the same homology class.} 
\label{regions} 
\end{figure}

Notice that the stability conditions can be much more complicated when more than two D-branes are present. For example, we can take three branes in such a way that every intersection is tachyon free (that can be achieved by taking, for example, three branes with wrapping numbers $(1,0)(1,0)(1,0)$, $(3,1)(3,1)(10,3)$ and $(2,1)(2,1)(2,1)$ in a square lattice). Now if we change the complex structure some tachyons will appear at the intersections between pair of branes. The moduli space is divided into zones where the stable system consists of a different numbers of branes. This process has been represented schematically in figure \ref{regions}.

\subsection{Orientifold}

The above constructions can be extended to cases with an orientifold plane (T-dual to Type I systems). Let us consider the orientifold plane lying on the $(1,0)(1,0)...$ cycle. The $\Omega$ acts together with an action in the space coordinates $R$ in the directions perpendicular on each $T^2$ to the orientifold plane. In order to get an invariant configuration one should consider pairs of branes related by $\Omega R$.

For example, take a two dimensional square torus. A D-brane wrapped on the $(n,m)$ cycle is taken to a brane wrapping $(n,-m)$. That means that the combined system has wrapping numbers: $2 (n,0)$. 

On a four dimensional torus, the  action can be guessed from the one cycle basis $(n,m) \rightarrow (n,-m)$: a brane wrapped on $(q_1,q_2,q_3,q_4)$ is taken to $(q_1,-q_2,-q_3,q_4)$. 

Something similar happens in the six dimensional torus where a brane wrapped on the $(q_1,q_2,q_3,q_4, q_5,q_6,q_7,q_8)$ is taken to another one wrapped on  $(q_1,-q_2,-q_3,q_4,$ $q_5,q_6,q_7,-q_8)$. 

In the T-dual picture this condition tells us that the bundle has vanishing first Chern number. We will come to this point later.

\section{Torons}

Brane at angles (A-branes) in toroidal compactifications are generically mapped by T-duality to branes wrapping the whole torus ( $B$ branes). These branes carry gauge bundles. The T-duality we are considering here is along one of the directions of each two dimensional torus.

Let us take a two dimensional torus with coordinates $0 \leq x_{1,2} \leq 2 \pi R$. Let us put a metric $g_{ij}$ and a $B_{12}$-field on it. These fields can be redefined as two complex scalars the complex structure $\tau = \tau_1 + \tau_2$ \footnote{Through this section we will follow the standard definitions of \cite{p98}.}:
\beq
\tau_1 = \frac{g_{12}}{g_{11}} \quad \tau_2 = \frac{\sqrt{det G}}{g_{11}}
\eeq
and the complexified Kahler form $\rho$: 
\beq
\rho_1 = b_{12} = \frac{B_{12} R^2}{\alpha'} \quad \rho_2 = \frac{A}{4 \pi \alpha'}
\eeq
where $A = 4 \pi^2 R^2 \sqrt{det G}$ is the area of the torus.

Now we perform a T-duality along the $x_1$ direction. T-duality interchanges the complex structure moduli with the Kahler one: $(\tau,\rho) \rightarrow (\rho,\tau)$. That means that the fields in the dual torus can be related to the old ones as:
\beqa
\rho_1' = \tau_1 & \longrightarrow & B_{12}' = \frac{\alpha'}{R'^2} \tau_1  \nonumber  \\
\rho_2' = \tau_2 & \longrightarrow & A' = 4 \pi^2 \alpha' \tau_2 \nonumber \\
& \tau_1' =  \frac{B_{12} R^2}{\alpha'} \nonumber &\\
& \tau_2' =  \frac{A}{4 \pi^2 \alpha'} &
\eeqa

Notice that if the B-field is absent in the brane at cycles side, the complex structure in the side with fluxes will be imaginary, i.e. a rectangular lattice. We will use this type of lattice along this paper, although the generalization to cases with a non-rectangular lattice is straightforward. 

A D-brane wrapping a straight line on the homology class $(m,n)$ forms an angle with the $x_1$-axis:
\beq
cotg \phi = \frac{m + n \tau_1}{n \tau_2}
\eeq
That means that a string ending on the brane has boundary conditions:
\beqa
\partial_{\sigma} (\sin{\phi} X_1 - \cos{\phi} X_2) = 0\nonumber \\
\partial_{\tau} (\cos{\phi} X_1 + \sin{\phi} X_2) = 0
\eeqa
T-duality takes this brane to a brane wrapping the whole torus with a magnetic field on it:
\beqa
\partial_{\sigma} X_1 - {\cal F}  \partial_{\tau} X_2  = 0\nonumber \\
{\cal F}  \partial_{\tau} X_1 +  \partial_{\sigma} X_2 = 0
\eeqa
where,
\beq
{\cal F} = cotg \phi = \frac{m}{n \tau_2} + \frac{\tau_1}{\tau_2}
\eeq

This field $\cal F$ is a combination of the $U(N)$ field strength:
\beq
F = 2 \pi \frac{m}{n A'} \id_n =  \frac{m}{2 \pi \alpha' n \tau_2} \id_n
\eeq
that has integer first Chern numbers, and a B-field. Then $\cal F$ can be written in the dual picture as:
\beq
{\cal F} = 2 \pi \alpha' F + \frac{B_{12}'}{\sqrt{G'}}
\label{relation}
\eeq

So a change in the complex structure on the branes at angles side is taken to a B-field flux on the other but preserving the integrability of the charges (homology or Chern classes on the other side). 

There are some subtleties that we will not considered here. If we take into account that in the T-duality process some of the branes are just in the direction of the T-duality, lower dimensional branes must be taken into account. For example, consider a D-brane wrapping a 1-cycle on a two dimensional torus. Now perform a T-duality along one direction. Generically, the D-brane direction does not coincide with the T-duality direction, so the dual system consists of a brane on the whole torus with some flux on it. However, if the two directions coincide the D-brane is taken to a lower dimensional brane located at one point on the dual torus. Mathematically that means that we are dealing with coherent sheaves rather than gauge bundles. We must also take into account that a negative rank is allowed, which makes sense in a K-theoretical way: a negative rank bundle corresponds to an antibrane with a bundle of positive rank. K-theory only takes into account the charges but not the moduli, for example, translations of the branes. In order to take all these features into account D-branes can be interpreted as objects in the derived category of coherent sheaves \cite{d00,k94,s99}. The classification of charges does not depend on the specific coherent sheaf or the corresponding object in the derived category because the K-theory of the category of coherent sheaves and of the derived category are identical. The Kontsevich proposal of mirror symmetry relates the derived category of coherent sheaves to the Fukaya category \footnote{A refined version of this is needed when the $B$-field is taken into account \cite{ko00}.} \cite{k94}. Our case constitutes a particular example of this duality, where some of the main issues can be analyzed in detail.

\subsection{Torons}

$U(N)$ gauge bundles are classified by a finite set of numbers, the Chern numbers. Once these numbers are specified there is a manifold of different connections not related by gauge transformation, the moduli. For example, take a $U(1)$ trivial bundle on a $T^2$, i.e. the first Chern class vanishes. The moduli consists of flat connections on the torus up to a gauge transformation. T-duality on the two directions tells us that the moduli is just the $T^2$. Chern classes specify completely the discontinuous part of the moduli  \cite{lpr}.

That is not the case for $SU(N)/Z_N$ gauge bundles \footnote{Or $SU(N)$ gauge fields transforming by conjugation.}. Consider a four dimensional torus. Because the $U(1)$ part is absent in this case all first Chern numbers vanish. However, there are $6N$ discrete choices that are not fixed (we will see that they are related to some integer numbers $n_{ij}$, where $n_{ij}= 0, \dots ,N-1$ and the indices label  1-cycles). So in this case, Chern numbers are not enough to classify the disconnected components, $n_{ij}$ numbers must be specified \cite{s82}. We will see how the $U(1)$ group of $U(N)$ is able to take into account these numbers.

Let us consider pure $U(N)$ Yang-Mills on a $2n$ dimensional torus \cite{gr971,gr972,cesar,thooft}. This torus can be constructed as a quotient of $R^{2n}$ by a lattice. A translation of an element of the lattice take a point to an equivalent one. Gauge fields at the translated point should be equivalent to the gauge field at the other point, i.e. they are related by a gauge transformation:
\beq
A^{j} (x^i + l^i) = \Omega_i A^{j} (x^i) 
\eeq
where $l_i$ is the period of $x_i$ and $\Omega_i$ is a $U(N)$ gauge transformation of the vector field in the adjoint:
\beq
\Omega_i A^{j} = \Omega_i (A^{j} - i \partial^j) \Omega_i^{-1}
\eeq  

We can split the $U(N)$ transformations into a $U(1)$ part and a $SU(N)$ contribution.  The $U(1)$ transition functions can be taken to:
\beq
\Omega_i = e^{2 \pi i \sum_j \frac{c_{ij} x_j}{l_j}}
\eeq
where $c_{ij}$ are some integer numbers that specify the bundle.

In order to define a good $U(N)$ bundle a closed path should give a trivial gauge transformation:
\beq
\Omega_i \Omega_j \Omega_i^{-1} \Omega_j^{-1} = \id 
\label{UN}
\eeq

Factorazing out the $U(1)$ part we get a condition for the $SU(N)$ bundle to be well defined into a $U(N)$ bundle:
\beq
\Omega_i \Omega_j \Omega_i^{-1} \Omega_j^{-1} = g 
\label{omega}
\eeq 
where g is an element of the center of $SU(N)$ (a phase times the unity). As gauge fields transform by conjugation, an element of the center of the group is acting as the unity.  An element of the center of $SU(N)$ can be written as a phase $e^{2 \pi i n_{ij}/N}$. The numbers $n_{ij}$ are integers from $i = 1, \dots, N$ that characterize different configurations on the torus. These numbers $n_{ij}$ are related to the $U(1)$ $c_{ij}$ numbers through the relation \ref{UN}:
\beq
n_{ij} = c_{ij} \ mod N
\eeq

Depending on the value of the second Chern number we can consider two cases \cite{gr971,gr972}:

\begin{itemize}

\item if the second Chern numbers of the $U(1)$ part are integer, i.e. 
\beq
\sum_{ijkl} \frac{\epsilon_{ijkl}c_{ij}c_{kl}}{8N} \in \IZ
\eeq
then we can find constant $SU(N)$ gauge transition functions \cite{thooft}. These functions do not have contribution to the Chern numbers but can compensate the $U(1)$ transformation  so the total $U(N)$ bundle is well defined. The $SU(N)$ transition functions can be written as constant matrices:
\beq
\Omega_i = P^{s_i} Q^{t_i} 
\eeq
where $PQ = QP e^{\frac{2 i \pi}{N}}$. Then, to find constant transition functions one should find $s_i$ and $t_i$ natural numbers which can reproduce the desired $n_{ij}$ numbers:
\beq
n_{ij} = s_i t_j - s_j t_i 
\eeq 

A necessary condition to find a solution to this equation is that the second Chern number is an integer. 

Notice that the $\Omega$'s form a projective representation of the $Z_N^{2n}$ (see relations (\ref{omega})) group. This representations are classified \cite{kar} by the group $H_2(G,U(1))$. In our case there are $2n N (2n-1)/2 $ representations \footnote{The  $H_2(G,U(1))$ of $G = Z_N^{2n}$ is just $Z_N^{2n(2n-1)/2}$.}. Notice that this is the number of $n_{ij}$'s, that are the possible gauge bundles for a fixed second Chern number in a $SU(N)/Z_N$ bundle.

\item In the case where the second Chern numbers are not integers we will not be able to find constant $SU(N)/Z_N$ transition functions. That means that there will be a $SU(N)$ flux giving contribution to higher Chern numbers. This flux will be responsible for the integral Chern numbers.

If this flux is diagonalized we can see that in this case the total flux can be decomposed into more than one  $U(1)$ fluxes. When translated into the dual system, this decomposition  means that we are dealing with several branes at different cycles.

\end{itemize}

Another way to think about it is from the $SU(N)$ point of view. An $SU(2)$ bundle on the four dimensional torus is specified by the $n_{ij}$ numbers and the $c_2$. If $c_2 = 0$ we can find configurations with $F_{ij}=0$ that have $n_{ij}$ boundary conditions \cite{thooft} based on a projective representation of some copies of the center of the group. When we try to make it compatible with a $U(1)$ field to be in a $U(N)$ group, some $U(1)$ flux appears because any $U(1)$ twist must carry a finite amount of action.

Let us first pay attention to the abelian part. In order to get a constant flux $F_{ij}$ we can choose gauge potentials:
\beq
A_{i} = \sum_j F_{ij} x_j
\eeq

These fields satisfy non-trivial boundary conditions. The values of these fields are discrete. That can be seen by transporting a unit charge around each plane:
\beq
F_{ij} l_i l_j = 2 \pi c_{ij}
\eeq
where the $l_i$ are the periods of the lattices defining the torus.
 
Inserting these fields into $U(N)$ matrices:
\beq
F_{ij} l_i l_j = 2 \pi \frac{c_{ij}}{N} \id_N
\eeq

This abelian part gives a contribution to the first Chern numbers:
\beq
c_1^{ij} = c_{ij}
\eeq

And to the second ones:
\beq
c_2^{ijkl} = \frac{c_{ij}c_{kl}}{N}
\eeq

Notice that these numbers can be fractional. Notice also that in this abelian case all Chern numbers can be obtained from the first Chern numbers.

When a system with several branes is  T-dualized the total gauge field is the sum of the contribution from each brane, that can be split into a $U(1)$ part plus $SU(N)$ contributions:
\beq
F = \sum_a \frac{n^a_{ij}}{N_a A_{ij}} \id_N^a
\eeq

This is a map from homology to K-theory charges on the torons side. Notice that there are $2^{d-1}$ Chern numbers that specify a $U(N)$ bundle in a $d$ dimensional torus. That coincides with the number of cycles we should consider in the dual case:

- in the two dimensional torus, there are two integers in the toron side: the rank and the first Chern number. In the branes at angles picture  there are two 1-cycles.

- In a four dimensional torus there are 8 integers: the rank, 6 first classes and the second one. In the other side there are 6 2-cycles, a 4-cycle and a 0-cycle.

- In the six dimensional torus there are 32 numbers: rank, 15 first Chern numbers, 15 second Chern numbers and 1 third Chern number. In the T-dual picture, there are all type of odd cycles: 6 5-cycles, 20 three-cycles and 6 1-cycles.

In the following sections we will clarify the relation between the configurations with brane at angles and the toron picture.

\subsection{Spectrum}

An approximative description of the low energy on a four dimensional torus with flux can be found in \cite{vb}, and generalized to higher dimensional tori in \cite{t}. The spectrum is obtained by considering small fluctuations of the gauge-fields around the vacuum with a constant field strength. The masses are obtained by keeping the quadratic terms in the action, which in the above cases is just the Yang-Mills action. 

To do so we separate the gauge potential into several pieces:
\beq
A_i = a_i^r T_r + b_i^{ab} e_{ab}
\eeq

The diagonal piece $a_i^r$ is proportional to the Cartan generators while the off-diagonal terms $ b_i^{ab}$ correspond to the other generators. In performing the T-duality the  $a_i^r$ fields will be related to the gauge fields and scalars in the adjoint coming from strings ending on the same brane. The non-diagonal fields are related to scalar fields and gauge fields coming from strings between two different sets of branes. In particular, it is possible that some of these fields will become tachyonic. Notice that physical fields should obey periodicity conditions to be well defined on the torus.

Thus, starting from the Yang-Mills functional one can obtain an approximated spectrum \cite{vb,t} \footnote{For the abelian case one can get the exact spectrum in $\alpha$' using open strings with a $U(1)$ field.}. In particular, between the less massive modes, the lightest off diagonal fields, $\phi^{ab}_k$, have a mass: $\frac{1}{2}(\sum_i f^{ab}_i - 2 f^{ab}_k)$, where $f^{ab}_i$ is the difference between the fields strengths in the $a$ and $b$ sectors in the $i$-th two dimensional torus.

From the spectrum is clear that the diagonal sector maps to strings joining the same brane and the non-diagonal sector maps to strings between different branes. These non-diagonal fields can become tachyonic in some regions of the flux moduli space. Chiral fermions can be obtained from an index theorem, just as in a KK reduction with non-vanishing fluxes. 

The correspondence between the two systems, brane at angles and Yang-Mills, is not exact due to the fact that we are not considering the whole Born-Infield action but just a first order approximation \cite{wt}. Notice that this discrepancy is the responsible of the mismatch of the scalar masses: in the brane at angles picture they are proportional to the angles while on the Yang-Mills side they are proportional to the fluxes, i.e. the tangents of the angles. 

This mismatch also appears in the supersymmetry conditions \footnote{See section 9 of \cite{t}.}. Let us consider a system of two branes on the four dimensional torus. Supersymmetry appears when one of the above scalar fields become massless, i.e. when $f_1$ is equal to $f_2$. That implies that the field strength is a solution to Hermitian Yang-Mills. However we know that Yang-Mills is not complete and some corrections are expected to appear. We will come later to this point when discussing supersymmetry conditions.

\subsection{Two dimensional torus}

Let us start with the two dimensional square torus of area $1$. In the brane at angle picture the brane system is specified by a set of integers $(n_a,m_a)$ for each brane. T-dualizing along the $[b]$ cycle direction, the brane  $(n_a,m_a)$ is mapped to a $U(n_a)$ gauge field with $c_{12} = m_a$. That corresponds to a field strength on the torus:
\beq
F_a = 2 \pi \frac{m_a}{n_a} \id_{n_a}
\eeq

Notice that $\frac{m_a}{n_a}$ is equal
 to  $tan \phi$, where $\phi$ is the angle of the brane with the x-axis.
This gauge bundle is classified by its first Chern number $c_1 = m_a$ (as $F_a$ is a constant field and the area is one the first Chern number is just the trace). A negative rank means that we are dealing with antibranes. As we have mentioned before, wrapping numbers are mapped to K-theory charges.  

The T-dual field strength of a system with several branes is just the sum of each bundle:
\beq
F = 2 \pi
\left(
\begin{array}{ccc}
\frac{m_1}{n_1} \id_{n_1} & & \\
& \ddots & \\
& & \frac{m_K}{n_K} \id_{n_K}
\end{array}
\right)
\eeq

Notice that the first Chern number is just the sum of the Chern number of each brane: $c_1 = \sum_1^K m_a$.

\subsection{Four dimensional torus}

Let us start from a $U(1)$ bundle. All Chern numbers can be obtained from the first Chern numbers. Let us first consider the case with only two non-vanishing fluxes, $F_{12}$ and $F_{34}$. The map can be easily be generalized from the two dimensional torus:
\beqa
N = q_1 \nonumber \\
c_{12} = q_3 \nonumber \\
c_{34} = q_2 
\eeqa

From these numbers, as we are in a $U(1)$ case, we can obtain the second Chern number, that will be directly identified with $q_4$:
\beq
c_2 = \frac{c_{12}c_{34}}{N} = \frac{q_2 q_3}{q_1} = q_4
\eeq

So we clearly see the relation between the Chern numbers in the $U(1)$ case and the factorizability of the 2-cycles. In this case, the brane system is specified by the numbers $q_i$ that can be obtained from $(n^1_a,m^1_a)$ and $(n^2_a,m^2_a)$. A negative $q_1$ bundle can be interpreted as an antibrane with positive $q_1$ and all the other charges reversed.

The map in the case of several branes states that the bundle is the sum of the bundle corresponding to  each brane. That means in terms of fluxes that:
\beq
F_{12} = 2 \pi
\left(
\begin{array}{ccc}
\frac{q_3^1}{q_1^1} \id_{q_1^1} & & \\
& \ddots & \\
& & \frac{q_3^K}{q_1^K} \id_{q_1^K}
\end{array}
\right) 
\eeq
and,
\beq
F_{34} = 2 \pi
\left(
\begin{array}{ccc}
\frac{q_2^1}{q_1^1} \id_{q_1^1} & & \\
& \ddots & \\
& & \frac{q_2^K}{q_1^K} \id_{q_1^K}
\end{array}
\right). 
\eeq

The second Chern number is the sum of the  second Chern numbers of each brane. In the language of \cite{gr971,gr972} that means that we should plug in a $SU(N)$ flux.

\subsection{Six dimensional torus}

Finally let us discuss the six dimensional torus. As in the previous case, we start with a diagonal $U(1)$ flux which corresponds in the dual case to one type of factorizable 3-cycle. The gauge bundle is a bundle with transition functions specified by three integers: $c_{12}$, $c_{34}$ and $c_{56}$, all the other transition numbers are zero. The map to the corresponding brane coordinates is the following:
\beqa
N = q_1 \nonumber \\
c_{12} = q_5 \nonumber \\
c_{34} = q_3 \nonumber \\
c_{56} = q_2 
\eeqa

As we are dealing with a $U(1)$ bundle, in the toron side all Chern numbers can be written in terms of the first Chern numbers. In the dual side, that means that there are relations between the wrapping numbers of the cycle that allow us to write every cycle in terms of the $q_1$, $q_2$, $q_3$ and $q_5$ wrapping numbers. 

The field strength in this case is: 
\beqa
F_{12} = 2 \pi  \frac{n_{12}}{N} \id_N = 2 \pi \frac{q_5}{q_1} \id_{q_1} \\
F_{34} = 2 \pi \frac{n_{34}}{N} \id_N = 2 \pi \frac{q_3}{q_1} \id_{q_1} \\
F_{56} = 2 \pi \frac{n_{56}}{N} \id_N = 2 \pi \frac{q_2}{q_1} \id_{q_1}
\eeqa

The second Chern numbers are identified (using the factorization conditions) with:
\beqa
\hat c_{12} = \frac{1}{(2 \pi)^2} tr (F_{34} F_{56}) = \frac{q_2 q_3}{q_1} = q_4 \nonumber \\
\hat c_{34} = \frac{1}{(2 \pi)^2} tr (F_{12} F_{56}) = \frac{q_2 q_5}{q_1} = q_6 \nonumber \\
\hat c_{56} = \frac{1}{(2 \pi)^2} tr (F_{12} F_{34}) = \frac{q_5 q_3}{q_1} = q_7 
\eeqa

The third Chern number can be written as: 
\beq
c_3 = \frac{1}{(2 \pi)^3} tr (F_{12} F_{34} F_{56}) = q_2 q_3 q_5 / (q_1)^2 = q_8
\eeq

So there is a one-to-one correspondence between Chern numbers and wrapping numbers. 

As the previous cases we can consider the dual system of a set of branes. The total dual bundle is just the sum of each $U(1)$ bundle.

\subsection{Tadpole cancellation conditions}

From the point of view of branes at angles the RR tadpole cancellation conditions state that the sum of the homology classes of all the branes is zero \cite{hepth,probes}:
\beq
\sum_a q_a^i = 0
\eeq

Each of these conditions corresponds to a RR field $C^i_{p+1}$. We can translate these conditions to gauge bundles using the above map:

\begin{itemize}
\item $\sum_a q_a^0 = 0$ is translated to the condition that the total rank should vanish, taking into account that a negative rank means that we are dealing with an antibrane with a positive rank.
\item $\sum_a q_a^i = 0$ is translated to the condition that the sum of all Chern numbers is zero for each direction.
\end{itemize}

These conditions should be imposed when the transverse space to the branes is compact, i.e. D8-branes wrapping 1-cycles on $T^2$, D7-branes on 2-cycles on $T^4$ and  D6-branes wrapping 3-cycles on $T^6$.

\subsection{Type I construction.}

In the brane at angles picture we have seen that, generically, in order to have a well defined orientifold action, branes should be put into pairs. For example, in the two dimensional case a brane on the $(n,m)$ is mapped by the orientifold action to another brane on the $(n,-m)$ cycle. 

Using the rules we have defined above we can see that we are dealing with the sum of two bundles with opposite first Chern number. That is the fluxes are of the form:
\beq
F = 2 \pi \sum_a \frac{c^a_{ij}}{N_a A_{ij}} 
\left(
\begin{array}{cc}
\id_N^a & 0 \\
0 & -\id_N^a
\end{array}
\right)
\eeq

So the first Chern numbers are zero. That does not mean that the other Chern numbers should vanish. In addition we can consider $N$ branes on an invariant cycle under the orientifold action, for example on the $(1,0)$ cycle in the two dimensional torus. As this set of branes is invariant under the orientifold action, it is mapped to itself. The action of the orientifold on the Chan-Paton matrices projects out some of the $U(N)$ modes to $SO(N)$ or $USp(N)$ groups. Tadpoles should take into account the charges of the orientifold planes. 

The brane at angles picture allows to understand some construction of stable non-BPS branes on Type I string theory \footnote{See, for example, \cite{lu01}.}.

\subsection{Symmetry groups}

Notice that in the brane at angles picture there is a symmetry that permutes all cycles preserving the intersection matrix. In the two dimensional torus this symmetry is the $SL(2,Z)$, in the four dimensional case is $Sp(6,Z)$ and in the six dimensional torus is the $Sp(20,Z)$. We can consider the subgroup that preserves the splitting of the higher dimensional torus into two dimensional ones. This subgroup is $(SL(2,Z))^n$ for a $T^{2n}$.

The action of this group on the cycles is mapped to an action on the bundles. Let us consider the two dimensional case. In this case there are two integers specifying the configuration $(n,m)$. Acting by an element of $SL(2,Z)$ this configuration is taken to:
\beq
\left(
\begin{array}{c}
n \\
m
\end{array}
\right)
\longrightarrow
\left(
\begin{array}{cc}
a & b \\
c & d
\end{array}
\right)
\left(
\begin{array}{c}
n \\
m
\end{array}
\right).
\eeq 

Notice that the complex structure also varies under this transformation. In the dual system this group acts on the Kahler moduli.

When translated into bundles this action takes a $U(n)$ gauge group with first Chern number $m$, to a gauge group $U(a n + b m)$ with first Chern number $c n + d m$. In particular, if the numbers $(n,m)$  are coprime the system can be taken to an equivalent one $(1,0)$. When translated into gauge bundles that means that for $U(n)$ with $c_1 = m$, with $n,m$ coprime,  we can always find an equivalent bundle with $U(1)$ and vanishing $c_1$. Even more, if $d$ is the greatest common divisor of these numbers, the system can always be taken to a $U(d)$ gauge group with vanishing $c_1$.

On higher dimensional tori, the intersection preserving group permutes the charges, i.e. the rank and Chern numbers in a complicated way. For example, the $(SL(2,Z))^2$ of $T^{4}$ changes the rank:
\beq
N \longrightarrow a_1 a_2 N + a_1 b_2 c_{34} + a_2 b_1 c_{12} +   b_1 b_2 c_2
\eeq

where $a_i, b_i, c_i, d_i$ are elements of the $(SL(2,Z))^2$ acting on the i-th two dimensional torus.

\section{Supersymmetry conditions}

We will analyze what are the conditions that a system of branes at angles should satisfy to preserve some supersymmetry. When the system is T-dualized supersymmetry conditions are taken to some conditions on the fluxes.

\subsection{From angles}

\subsection{Two dimensional torus}

\begin{figure}
\centering
\epsfxsize=3in
\hspace*{0in}\vspace*{.2in}
\epsffile{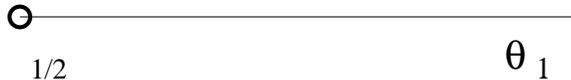}
\caption{\small Angle parameter space for a system of two branes wrapping 1-cycles on $T^2$.} 
\label{T2} 
\end{figure}

Let us consider a pair of D-branes wrapping 1-cycles. As we have discussed previously there is a tachyon at each intersection with a mass square proportional to the angle. The system is non-supersymmetric generically. Only when the angle is zero, i.e. the two branes are parallel, the system is supersymmetric. We can plot the angle between these two branes (see figure \ref{T2}). The supersymmetric point preserves one half of the supersymmetries.

This condition generalizes for several branes: all must be parallel to be supersymmetric. When this condition is translated into the dual system of fluxes,
\beq
\forall a, \theta_a = \alpha \Leftrightarrow  {\cal F} = \mu \id
\eeq
where $\theta_a$ is the angle of the $a$ brane with respect to a fixed line, and $\alpha$ and $\mu$ are constants. We have seen in section 3 that the B-dependence in this equation can be eliminated by identifying the dependence in the complex structure in the dual model.

From the equation above we can see  that the parallelism of all branes is equivalent in the dual picture to a flux that satisfies the Hermitian Yang-Mills equations. A solution to this equation is equivalent, by the Donaldson-Uhlenbeck-Yau correspondence, to a stable bundle.

\subsection{Four dimensional torus}

\begin{figure}
\centering
\epsfxsize=3in
\hspace*{0in}\vspace*{.2in}
\epsffile{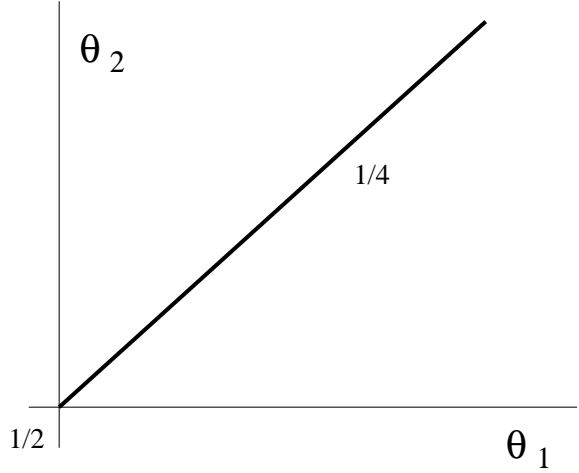}
\caption{\small Angle parameter space for a system of two branes wrapping 2-cycles on $T^4$.} 
\label{T4} 
\end{figure}

In this case there are two angles to be taken into account. One quarter of the supersymmetry is preserved when the absolute value of these angles is equal (see figure \ref{T4}):
\beq
|\theta_1| = |\theta_2|
\eeq

The extremal case where both angles vanish preserves one half of the supersymmetries.

Let us consider both angles positive. In the general case, when more than two branes are considered, the angles in each two dimensional torus for every brane must be equal. Supersymmetry condition can be rewritten as:
\beq
\theta^a_1 - \theta^a_2 = \alpha
\eeq
for each brane.

Taking the tangent of this equation and  using the map given above, this condition is translated to:
\beq
{\cal F}_{12} + {\cal F}_{34} = \mu (\id - {\cal F}_{12}{\cal F}_{34})
\label{susy4}
\eeq

Higher powers of the field strength correspond to $\alpha'$ corrections. That means that for $\alpha'$ close to zero the equation reproduces ordinary Hermitian-Yang Mills.
 
Taking into account the combinatorial factors and that the Kahler form split into the Kahler form for each two torus we can see that the  equation \ref{susy4} is a special case of the non-abelian MMMS equation \cite{mmms99}:
\beq
J \wedge {\cal F} = \mu (\frac{1}{2}J \wedge J - \frac{1}{2} {\cal F} \wedge {\cal F})
\eeq

\subsection{Six dimensional torus}

Finally we will consider the six dimensional torus. Now there are four scalars that can become tachyonic depending on the angles. These scalars have a mass that can be expressed in terms of the angles at the intersection (see equations (\ref{tachyons})). When one of these scalars fields is massless there is a fermion boson degeneracy that indicates that $1/8$ supersymmetries are preserved. If there are two of these scalars that are massless a higher amount of supersymmetry is preserved ($1/4$).

The angle parameter space can be represented as a tetrahedron. On the walls of the tetrahedron one of the supersymmetries is preserved. Outside the tetrahedron one of the above scalars become tachyonic. That indicates that the system is not stable and that there is another system, wrapping the same cycles, that lowers the energy. Inside the tetrahedron the system is not tachyonic, so there is not any other configuration in the same homology class with a lower energy. We will see the meaning of these conditions in the following section.

\begin{figure}
\centering
\epsfxsize=3in
\hspace*{0in}\vspace*{.2in}
\epsffile{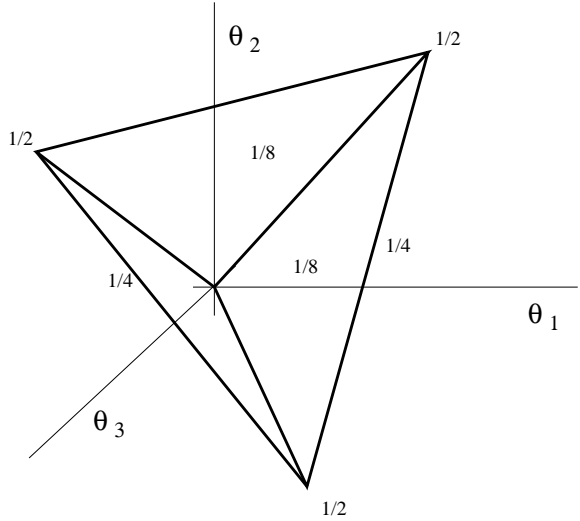}
\caption{\small Angle parameter space for a system of two branes wrapping 3-cycles on $T^6$.} 
\label{T6} 
\end{figure}

On each wall a different supersymmetry is preserved. That is the reason why on the intersection of two walls, the edges of the tetrahedron, there is a double number of supersymmetries preserved. On the vertices of the tetrahedron half of the supersymmetries are preserved.

The above conditions can easily be mapped to torons satisfying the MMMS equations. There is one condition for each tetrahedron wall. Let us take one of the walls: 
\beq
m^2_4 \alpha' = 1 - \frac{1}{2}( \theta_1 +\theta_2 +\theta_3) = 0
\eeq

Take an axis, the tangent and expanding it, one obtains the following equation:
\beq
{\cal F}_{12} + {\cal F}_{34} + {\cal F}_{56} - {\cal F}_{12}{\cal F}_{34}{\cal F}_{56} = \mu (\id - ({\cal F}_{12}{\cal F}_{34} + {\cal F}_{12}{\cal F}_{56} + {\cal F}_{56}{\cal F}_{34}))
\eeq

The conditions for the other faces are quite similar just taking into account the change of signs.

As in the four dimensional case this equation is a special case of the  non-abelian MMMS equation \cite{mmms99}:
\beq
\frac{1}{2} J \wedge J \wedge {\cal F} -  \frac{1}{3!} {\cal F} \wedge {\cal F} \wedge {\cal F} = \mu (\frac{1}{3!} J \wedge J \wedge J -  \frac{1}{2} J \wedge{\cal F} \wedge {\cal F} )
\eeq

To appreciate the difference between usual Yang-Mills and the T-dual system of branes at angles we plot the supersymmetry conditions in the Yang Mills case \cite{t}:
\beqa
- tan \theta_1 + tan \theta_2 + tan \theta_3 = 0 \nonumber \\
tan \theta_1 - tan \theta_2 + tan \theta_3  = 0 \nonumber \\
tan \theta_1 + tan \theta_2 - tan \theta_3  = 0
\label{YM}
\eeqa

The equation corresponding to the last wall of the tetrahedron is not present because it corresponds to a string excitation.

Notice the resemblance between these equations and the tetrahedron walls specified by (\ref{tachyons}). The supersymmetry conditions obtained in the Yang-Mills case are related to the tangent of the angle instead to the angles expected from the branes at angles picture. Both conditions coincide in the limit where at least one of the angles vanishes. If only one of the angles vanish the other two must be equal, i.e. we are on the edges of the tetrahedron where $1/4$ of the supersymmetry are preserved. Thus, supersymmetry from Yang-Mills and from angles coincide when more at least $1/4$ of the supersymmetries are preserved. Of course, near the origin, with $1/2$ of the supersymmetry preserved, these supersymmetry conditions get closer and closer. This seems to indicate that Yang-Mills can only reproduce the expected results from the dual side when a more than one supersymmetry is present.

\begin{figure}
\centering
\epsfxsize=3in
\hspace*{0in}\vspace*{.2in}
\epsffile{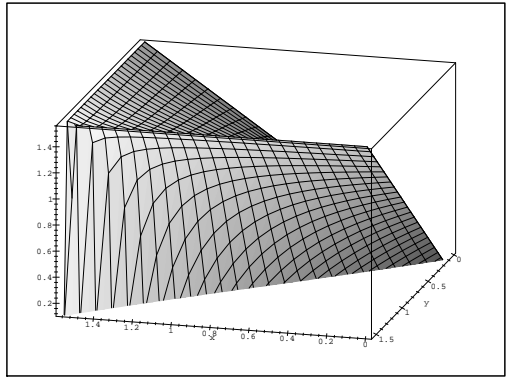}
\caption{\small Angle parameter space on $T^6$ from Yang Mills perspective. The expected tetrahedron is deformed far away from the origin.} 
\label{YM1} 
\end{figure}

\begin{figure}
\centering
\epsfxsize=3in
\hspace*{0in}\vspace*{.2in}
\epsffile{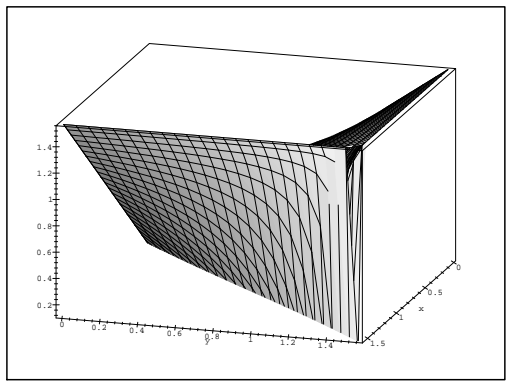}
\caption{\small Another view of the deformed tetrahedron.} 
\label{YM2} 
\end{figure}

\begin{figure}
\centering
\epsfxsize=3in
\hspace*{0in}\vspace*{.2in}
\epsffile{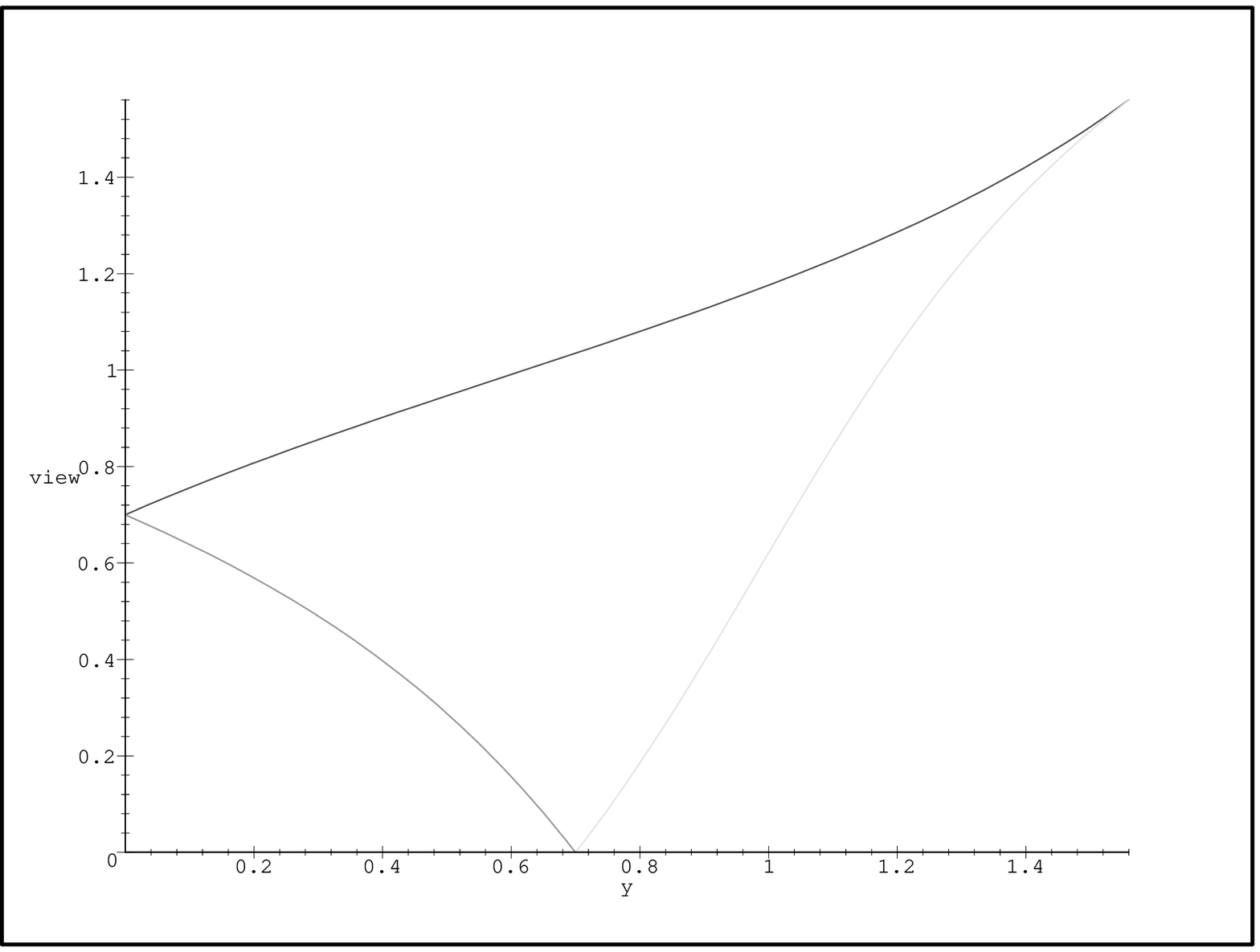}
\caption{\small Section of the deformed tetrahedron.} 
\label{section} 
\end{figure}

See the figures \ref{YM1} and  \ref{YM2} where the modified tetrahedron (\ref{YM}) is plotted. The figure is not longer a tetrahedron. Near the origin the difference cannot be appreciated, but  far from the origin the walls join together into a point. Regions that preserve one quarter of the supersymmetries are not modified as expected from lower dimensional tori expectations. We can see from the figure \ref{section} that the region inside the tetrahedron is deformed. The faces, $1/8$ SUSY preserving configurations, are not longer planes in the $\theta$-space.

\section{Stability regions}

\subsection{Stability and tachyons in the branes at angles side}

As we have mentioned before stability is related to the absence of tachyons \footnote{A different approach can be taken by studying the potential between pairs of branes (see, for instance,\cite{potential}).}. Tachyons can appear at every intersection. In the two and four dimensional tori supersymmetry is the only chance for a system to be stable. That is not the case in the six dimensional torus: there is a region, inside the tetrahedron where a pair of branes can be stable and not supersymmetric. When more than two branes are present in the system, stability should be checked from the angles between every pair of branes, i.e. every angle must be inside the tetrahedron.  

From a more geometric point of view  stability is related to minimal volume configurations. A brane that lowers its volume is an straight line in the two dimensional torus, a special Lagrangian manifold in the four and six dimensional tori \cite{bbh00,km99}. 

Consider a pair of branes wrapping submanifolds of minimal volume in different homology classes, these classes being the charges of the branes. Charges must be conserved, so we expect them to decay into a minimizing volume manifold in the same homology cycle that the sum of the homology cycles of the original brane. That is always the case in the two and four dimensional tori. In the four dimensional tori there is a locus in the space of complex structures where the sum of the cycles is semistable.

A more complicate structure appears in the six dimensional torus where for some values of the complex structure parameters the configuration that minimizes the volume is the sum of the two cycles. As an illustrative example, consider a pair of D6-branes on a six dimensional torus. Moreover we will consider a factorized torus with complex structure $\tau_i$ and the pair of branes wrapping the $(1,0)(1,0)(1,0)$ and $(1,1)(1,1)(1,1)$ cycles. The reduced complex moduli space consists of three copies of the upper plane $(\tau_1,\tau_2,\tau_3)$. The map from the reduced complex moduli to the angles space is:
\beq
\cos \theta_i = \frac{1+ Re(\tau_i)}{\sqrt{1+2 Re(\tau_i) + |\tau_i|^2 }}
\label{cos}
\eeq

A wall in the angle space is taken by the inverse map of (\ref{cos}) to a wall in the complex structure space, because the above equation is a real equation so the subspace that satisfies it has real codimension one.

\subsection{Stability and tachyons in torons side}

Stability conditions in the branes at angles side are taken by T-duality to some stability conditions on the bundles on the other side. For the two dimensional case the notion of stability from the branes at angles side is equivalent to $\mu$-stability. That is not the case for higher dimensional tori. Notice that the stability condition inherited from the branes at angles side is analogous to the $\Pi$-stability of \cite{dfr00}.

Let us define $\mu$-stability to compare both notions from our previous results. The degree of a bundle $E$ in a Kahler manifold with a Kahler form $J$ is:
\beq
deg_J (E) = \int c_1 \wedge J^{d-1}
\eeq
where $d$ is the complex dimension of the manifold. From the degree the slope can be defined as:
\beq
\mu_J (E) = \frac{1}{N} deg_J (E)
\eeq

A bundle $E$ is said to be $\mu$-stable if for every subbundle $E'$ the slope of the subbundle is less than the slope of the bundle. The bundle is semistable if the inequality is saturated.

Let us consider the easiest example, the two dimensional torus. Take one brane wrapping the 1-cycle $(n,m)$. The dual system is a $U(n)$ gauge group with $c_1 = m$. The degree is just the $c_1$ and the slope:
\beq
\mu = \frac{1}{N}c_1 = \frac{m}{n}
\eeq

Notice that the slope is the slope of the brane respect to the $x_1$-cycle, $tan \phi$ in our notation, in the dual system.

\begin{figure}
\centering
\epsfxsize=3in
\hspace*{0in}\vspace*{.2in}
\epsffile{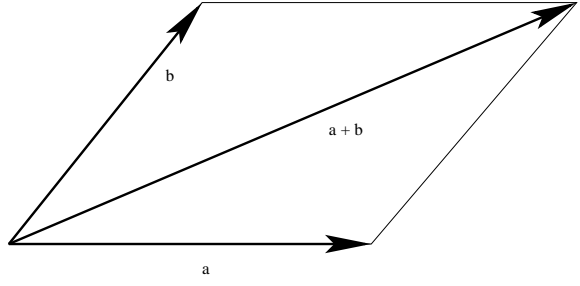}
\caption{\small The total bundle $a+b$ is composed of two  subbundles $a$ and $b$. The brane $a + b$ is not stable because there is a subbundle $b$ with higher slope. }
\label{mu} 
\end{figure}

If there are several branes, the T-dual system is the sum of the bundles of each one separately. Then the total bundle has $c_1 = \sum_a^K m_a$ and total rank $N = \sum_a^K n_a$. The degree is then $\sum_a^K m_a$ and the slope:
\beq
\mu = \frac{1}{N}c_1 = \frac{\sum_a^K m_a}{\sum_a^K n_a}
\eeq

We can see that the notion of stability we have defined from T-dual picture coincides with the $\mu$-stability. That is very easy to see because if the branes are not parallel we can find a subbundle with higher slope than the total (see figure \ref{mu}). Notice that when the system is stable is indeed semistable (supersymmetric). Notice also that the slope is the real slope in the T-dual picture. 

We know that this system is unstable (generically if the branes are not parallel) and it is going to decay to a brane wrapping a cycle $n = \sum_a^K n_a$ and $m = \sum_a^K m_a$. T-duality takes the new stable system to a $U(n)$ gauge group with $c_1 = m$. The slope of this bundle is:
\beq
\mu = \frac{1}{N}c_1 = \frac{m}{n}
\eeq

Notice that the presence of a B-field on the two dimensional torus changes reflects that we have changed the complex structure in the branes at angles picture. So the angle has changed, but not the slope $\mu$, because in the definition of it the presence of the B-field is not taken into account. The slope $\mu$ and the slope of the dual picture $tan \phi$ do not coincide. However, there is a correspondence between them as we have explained in section 3 \ref{relation}:
\beq
tan \phi = \mu + B_{12}'
\eeq

For a higher dimensional torus with fluxes in orthogonal two torus  we have seen that the flux can be obtained from the $n_{ij}$ numbers and the total rank. Taking a squared lattice, the degree of the bundle can be easily obtained:
\beq
deg_J (E) = \sum_{ij} n_{ij}
\eeq

The slope of the bundle $E$ is:
\beq
\mu_J (E) = \frac{1}{N} deg_J (E) = \frac{1}{N}\sum_{ij} n_{ij}
\eeq
 
For instance, take the four dimensional torus that we have described above. The degree of the bundle is:
\beq
deg_J (E) = n_{12} + n_{34} = q_3 + q_2
\eeq
and the slope:
\beq
\mu_J (E) = \frac{q_3 + q_2}{q_1} = \frac{n_1 m_2 + n_2 m_1}{n_1 n_2} = \frac{m_2}{n_2} + \frac{m_1}{n_1} 
\eeq
 
The total slope is the sum of the slopes on each two torus. 

Consider a pair of branes with a bundle $E_a$ and $E_b$ with positive rank (we are not considering antibranes here). The total bundle $E_a \oplus E_b$ has a slope:
\beq
\mu(E_a \oplus E_b) (N_a + N_b) = \mu(E_a) N_a + \mu(E_b) N_b
\eeq 

The only chance for the total bundle to be semistable
\beqa
\mu(E_a \oplus E_b) \ge \mu(E_a) \nonumber \\ 
\mu(E_a \oplus E_b) \ge \mu(E_b) 
\eeqa 
is to have $\mu(E_a) = \mu(E_b) = \mu(E_a \oplus E_b)$. In the four dimensional factorizable case this implies that:
\beq
tan \phi_1^a + tan \phi_2^a = tan \phi_1^b + tan \phi_2^b
\eeq

That is not the condition expected from the branes at angles side:
\beq
tan (\phi_1^a + \phi_2^a) = tan (\phi_1^b + \phi_2^b)
\eeq

Notice that 'negative rank bundles' are interpreted as antibranes. That means that the $\mu$-stability condition must be completed to admit antibranes  \footnote{This fact is taken into account in the stability conditions proposed in relation to triples (brane, antibrane, tachyon) have been mentioned in \cite{opw00} in relation to \cite{bg94}. A similar system has been studied in \cite{tatar}.}. Notice also that branes at angles stability takes into account branes and antibranes (from the branes at angles point of view this transition is just a change in the angle from $0$ to $\pi$).

In higher dimensional tori $\mu$-stability and $\Pi$-stability (the stability condition from the T-dual picture) do not coincide. Some corrections are expected to the definition of slope coming from higher order Chern classes due to string corrections and from the presence of the B-field. All this corrections are taken into account in the brane at angle picture.

\bigskip

\bigskip

\bigskip

\centerline{\bf Acknowledgements}

I am grateful to Luis Alvarez-Consul, Daniel Cremades, David Garc\'{\i}a, Luis Ib\'a\~{n}ez, Fernando Marchesano and specially to Angel Uranga for useful discussions. The research of R.R. was supported by the MECD through a FPU grant.


\end{document}